\documentstyle[12pt] {article}
\def\be{\begin{equation}}
\def\ee{\end{equation}}
\def\bea{\begin{eqnarray}}
\def\eea{\end{eqnarray}}
\def\pd{\partial}

\def\eps{\epsilon}
\def\vmu{\vec\mu}
\def\vnu{\vec\nu}
\def\vrho{\vec\rho}
\def\vkap{\vec\kappa}
\def\vkp{\vec\kappa'}
\def\vkpp{\vec\kappa''}

\def\vsig{\vec\sigma}
\def\vtau{\vec\tau}
\begin{document}

\begin{flushright} DTP/00/39\\[1cm]
\end{flushright}

\begin{center}
\large{\bf  Equivalence Theorem for Lagrangians in Different Dimensions}\\[5mm]
\large L. M. Baker$\footnote{e-mail: l.m.baker@durham.ac.uk}$\\[5mm]
Department of Mathematical Sciences,\\
University of Durham,\\
South Road,\\
Durham, England, DH1 3LE\\[10mm]
\end{center}

\begin{abstract}
A proof is given for the observation that the equations of motion for the companion Lagrangian without a square root, subject to some constraints, just reduce to the equations of motion for the companion Lagrangian with a square root in one less dimension.  The companion Lagrangian is just an extension of the Klein-Gordon Lagrangian to more fields in order to provide a field description for strings and branes.

\end{abstract}

\section{Introduction}
In \cite{me1},\cite{me2} a proposal was made to extend the wave-particle duality where a particle can be described by the Lagrangian ${\cal L}_1$ or terms of a Klein-Gordon field ${\cal L}_2$
\be
{\cal L}_1\,=\,\sqrt{\left(\frac{\pd X^\mu}{\pd\tau}\right)^2},\qquad{\cal L}_2\,=\,\ \frac{1}{2}{\left(\frac{\pd \phi}{\pd x_\mu}\right)^2}\label{one}
\ee
 to extended objects so that strings and branes with Lagrangian ${\cal L}_3$ could also be described by a theory with Lagrangian ${\cal L}_4$.
\be
{\cal L}_3\,=\,\sqrt{{\mathrm det}\left|\frac{\pd X^\mu}{\pd\sigma^i}\frac{\pd X^\mu}
{\pd\sigma^j}\right|},\qquad{\cal L}_4={\mathrm det}\left|\frac{\pd\phi^i}{\pd x_\mu}\frac{\pd\phi^j}{\pd x_\mu}\right|\label{two}
\ee
The number of fields $\phi^i$ is the same as the number of world-sheet co-ordinates $\sigma^i$. The new Lagrangian will be referred to as the companion Lagrangian. 
Although ${\cal L}_4$ is the natural extension of the Klein-Gordon Lagrangian, it would be preferable to consider $\sqrt{\cal L}_4$ since this possesses general covariance. The following result allows us to link these two Lagrangians.

In \cite{me2} it was noted that the equations of motion for the companion Lagrangian without a square root, when subjected to some constraints, reduce to the equations of motion for the companion Lagrangian with a square root in one dimension less. Computer calculations had verified this in several cases and an analytic proof was given for the case of two fields. Here we give an analytic proof of this observation for any number of fields $n$ in any number of dimensions $d$ where $d>n$. This result can most easily be checked for the case of one field.

\section{Conventions and Notation}
Partial derivatives are denoted by
\be
\frac{\pd\phi^i}{\pd x_\mu}=\phi^i_\mu,\qquad \frac{\pd^2\phi^i}{\pd x_\mu \pd x_\nu}=\phi^i_{\mu\nu}
\ee
Totally antisymmetric tensors $\eps_{\nu_1\nu_2\ldots\nu_d}$ are used throughout the proof with the convention $\eps_{12\ldots d}=+1$. When indices have an arrow above them then they represent several indices. They can be thought of as vectors with several components.\\
$\vmu,\vnu,\vrho,\vsig$ each have $(n-1)$ components. For example $\vmu$ denotes $\{\mu_2, \mu_3,\ldots,\mu_n\}$. \\
$\vtau, \vkap$ each have $(d-n)$ components. For example $\vkap$ denotes $\{\kappa_1, \kappa_2,\ldots,\kappa_{d-n}\}$.\\
$\vkp$ denotes $\{\kappa_2,\kappa_3,\ldots,\kappa_{d-n}\}$ and $\vkpp$ denotes $\{\kappa_3,\ldots,\kappa_{d-n}\}$.\\
For the product of $(n-1)$ fields we use the notation $\Phi_{\vnu}=\phi^2_{\nu_2}\phi^3_{\nu_3}\ldots\phi^n_{\nu_n}$.\\[2mm]
A useful identity which will be used later on is
\bea
\eps_{\mu\nu_2\nu_3\ldots\nu_d}\eps_{\rho_1\rho_2\ldots\rho_d}=\eps_{\rho_1\nu_2\nu_3\ldots\nu_d}\eps_{\mu\rho_2\ldots\rho_d}+\eps_{\rho_2\nu_2\nu_3\ldots\nu_d}\eps_{\rho_1\mu\rho_3\ldots\rho_d}+\ldots\nonumber\\
\ldots+\eps_{\rho_d\nu_2\nu_3\ldots\nu_d}\eps_{\rho_1\rho_2\ldots\rho_{d-1}\mu}\label{epid}
\eea
It amounts to swapping the index $\mu$ from the first epsilon with each index from the second epsilon.
\section{Lagrangians and Equations of Motion}

Consider the Lagrangian for $n$ fields $\phi^i$ in $d$ space-time dimensions $x^\mu$ which does not involve a square root.
\be
{\cal L}= {\mathrm det}\left|\frac{\pd\phi^i}{\pd x_\mu}\frac{\pd\phi^j}{\pd x_\mu}\right|
\ee
The equations of motion for this Lagrangian are
\be
\frac{\pd^2 {\cal L}}{\pd \phi^i_\mu \pd \phi^j_\nu} \phi^j_{\mu \nu} = 0 \label{eomnsq}
\ee
These determinantal Lagrangians can be written as the sum of squares of Jacobians. The Jacobians will be denoted as
\be 
J_{\vkap}=J_{\kappa_1\kappa_2\ldots\kappa_{d-n}}=\eps_{\kappa_1\kappa_2\ldots\kappa_{d-n}\nu_1\nu_2\ldots\nu_n}\phi^1_{\nu_1}\phi^2_{\nu_2}\ldots\phi^n_{\nu_n}
\ee
For the square root case the Lagrangian is
\be
{\cal L}= \sqrt{{\mathrm det}\left|\frac{\pd\phi^i}{\pd x_\mu}\frac{\pd\phi^j}{\pd x_\mu}\right|}=\sqrt{\frac{1}{(d-n)!}J_{\vkap} J_{\vkap}}\label{lagsq}
\ee
The equations of motion for this can be written as
\be
J_{\mu\vkp} J_{\nu\vkp} \phi^i_{\mu \nu}=0\label{eomsq}
\ee

\section{The Constraints}
The equations of motion for the non-square root case will be subject to the following constraints.
\be
\frac{\pd{\cal L}}{\pd \phi^i_\mu} \phi^i_{\mu \nu}=0. \label{const}
\ee
There is no summation over the index $i$.
The Lagrangian must also vanish.

The idea is to reduce the number of dimensions from $d$ to $d-1$. The constraints (\ref{const}) can be used to eliminate all second derivatives of the fields which involve a partial derivative with respect to $x_d$, the $d$th dimension. i.e From the constraints
\bea
\phi^i_{d \beta} = - \frac{\frac{\pd {\cal L}}{\pd \phi^i_\alpha}}{\frac{\pd {\cal L}}{\pd \phi^i_d}} \phi^i_{\alpha \beta}, \qquad
\phi^i_{dd} = - \frac{\frac{\pd {\cal L}}{\pd \phi^i_\alpha}\frac{\pd {\cal L}}{\pd \phi^i_\beta}}{\left(\frac{\pd {\cal L}}{\pd \phi^i_d}\right)^2} \phi^i_{\alpha \beta}
\eea
Putting these into the equations of motion (\ref{eomnsq}) we have:
\bea
\sum_{j=1}^n\frac{1}{\left(\frac{\pd {\cal L}}{\pd\phi^j_d}\right)^2}\Biggl[\left(\frac{\pd {\cal L}}{\pd\phi^j_d}\right)^2 \frac{\pd^2 {\cal L}}{\pd\phi^i_\alpha \pd\phi^j_\beta}-\frac{\pd {\cal L}}{\pd \phi^j_d} \frac{\pd {\cal L}}{\pd \phi^j_\beta}\frac{\pd^2 {\cal L}}{\pd\phi^i_\alpha \pd\phi^j_d}
-\frac{\pd {\cal L}}{\pd \phi^j_d} \frac{\pd {\cal L}}{\pd \phi^j_\alpha}\frac{\pd^2 {\cal L}}{\pd\phi^i_\beta \pd\phi^j_d}\nonumber\\+\frac{\pd {\cal L}}{\pd\phi^j_\alpha}\frac{\pd {\cal L}}{\pd\phi^j_\beta} \frac{\pd^2 {\cal L}}{\pd\phi^i_d \pd\phi^j_d}\Biggr]\phi^j_{\alpha \beta}=0 \label{pr1}
\eea
The indices $\alpha,\beta=1,2,\ldots,(d-1)$ throughout the paper.

\section{The Proof}
For the moment we shall consider the equation of motion with respect to field $\phi=\phi^1$ and are only looking at the component which involves the terms $\phi_{\alpha \beta}$. The other components will work in the same way. 
Using the definition of the Lagrangian which involves the Jacobians then we can write
\be
{\cal L} = \frac{1}{(d-n)!}\phi_\nu \phi_\rho B_{\nu\rho} \qquad {\mathrm where} \qquad  B_{\nu\rho}=\eps_{\nu\vkap\vnu}\eps_{\rho\vkap\vrho} \Phi_{\vnu} \Phi_{\vrho}
\ee
so the numerator of the coefficient of $\phi_{\alpha \beta}$ in (\ref{pr1}) becomes
\be
[B_{\mu d} (B_{\nu d} B_{\alpha \beta}-B_{\nu \beta} B_{\alpha d})+B_{\nu \alpha} (B_{\mu \beta} B_{dd}-B_{\mu d} B_{\beta d})]\phi_\mu \phi_\nu\label{pr2}
\ee
Now, 
\bea
B_{\nu d}B_{\alpha \beta}-B_{\nu \beta}B_{\alpha d}&=&[\eps_{\nu\vkap\vmu}\eps_{d\vkap\vnu}\eps_{\alpha\vtau\vrho}\eps_{\beta\vtau\vsig}-\eps_{\nu\vkap\vmu}\eps_{\beta\vkap\vnu}\eps_{\alpha\vtau\vrho}\eps_{d\vtau\vsig}]\ \Phi_{\vmu}\Phi_{\vnu}\Phi_{\vrho}\Phi_{\vsig}\nonumber\\
&=&\eps_{\nu\vkap\vmu}\eps_{\alpha\vtau\vrho}[\eps_{d\vkap\vnu}\eps_{\beta\vtau\vsig}-\eps_{\beta\vkap\vnu}\eps_{d\vtau\vsig}]\Phi_{\vmu}\Phi_{\vnu}\Phi_{\vrho}\Phi_{\vsig}\label{pr5}
\eea
Using the epsilon identity (\ref{epid}) to move the index $\beta$ around
\bea
\eps_{d\vkap\vnu}\eps_{\beta\vtau\vsig}=\eps_{\beta\vkap\vnu}\eps_{d\vtau\vsig}+\eps_{d\beta\kappa_2\ldots\kappa_r\vnu}\eps_{\kappa_1\vtau\vsig}+\ldots+\eps_{d\kappa_1\ldots\kappa_{r-1}\beta\vnu}\eps_{\kappa_r\vtau\vsig}+\nonumber\\\eps_{d\vkap\beta\nu_3\ldots\nu_n}\eps_{\nu_2\vtau\vsig}+\ldots+\eps_{d\vkap\nu_2\ldots\nu_{n-1}\beta}\eps_{\nu_n\vtau\vsig}\label{pr3}
\eea
The first term on the right hand side is just the other term in expression (\ref{pr5}). The last $(n-1)$ terms will all vanish due to symmetry conditions. This only leaves the middle terms. But by relabelling
\bea
\eps_{\nu \kappa_1\ldots\kappa_r\vmu}\eps_{d\kappa_1\ldots\beta\ldots\kappa_r\vnu}\eps_{\kappa_i\vtau\vsig}&=&\eps_{\nu \kappa_i\ldots\kappa_1\ldots\kappa_r\vmu}\eps_{d\kappa_i\ldots\beta\ldots\kappa_r\vnu}\eps_{\kappa_1\vtau\vsig}\nonumber\\
&=&\eps_{\nu\vkap\vmu}\eps_{d\beta\kappa_2\ldots\kappa_r\vnu}\eps_{\kappa_1\vtau\vsig}
\eea
There are $r=d-n$ of these terms. Therefore,
\be
B_{\nu d}B_{\alpha \beta}-B_{\nu \beta}B_{\alpha d}=r\eps_{\nu\vkap\vmu}\eps_{\alpha\vtau\vrho} \eps_{d\beta\vkp\vnu}\eps_{\kappa_1\vtau\vsig}\Phi_{\vmu}\Phi_{\vnu}\Phi_{\vrho}\Phi_{\vsig}
\ee  
Now as in (\ref{pr3}), using identity (\ref{epid}) to swap subscript $\kappa_1$
\bea
\eps_{\nu\vkap\vmu}\eps_{\alpha\vtau\vrho}=\eps_{\nu\alpha\kappa_2\ldots\kappa_r\vmu}\eps_{\kappa_1\vtau\vrho}+\eps_{\nu\tau_1\kappa_2\ldots\kappa_r\vmu}\eps_{\alpha\kappa_1\tau_2\ldots\tau_r\vrho}+\ldots+\eps_{\nu\tau_r\kappa_2\ldots\kappa_r\vmu}\eps_{\alpha\tau_1\ldots\tau_{r-1}\kappa_1\vrho}\nonumber\\+\eps_{\nu\rho_2\kappa_2\ldots\kappa_r\vmu}\eps_{\alpha\vtau\kappa_1\rho_3\ldots\rho_n}+\ldots+\eps_{\nu\rho_n\kappa_2\ldots\kappa_r\vmu}\eps_{\alpha\vtau\rho_2\ldots\rho_{n-1}\kappa_1}
\eea
And by relabelling indices and using the antisymmetric property of the epsilons
\be
\eps_{\nu\tau_i\kappa_2\ldots\kappa_n\vmu}\eps_{\alpha\tau_1\ldots\kappa_1\ldots\tau_r\vrho}\eps_{\kappa_1\vtau\vsig}=-\eps_{\nu\vkap\vmu}\eps_{\alpha\vtau\vrho}\eps_{\kappa_1\vtau\vsig}
\ee
so,
\be
(1+r) \eps_{\nu\vkap\vmu}\eps_{\alpha\vtau\vrho}\eps_{\kappa_1\vtau\vsig}=\eps_{\nu\alpha\vkp\vmu} \eps_{\kappa_1\vtau\vrho}\eps_{\kappa_1\vtau\vsig}
\ee
which gives
\be
B_{\nu d}B_{\alpha \beta}-B_{\nu \beta}B_{\alpha d}=\frac{r}{r+1} B_{\kappa\kappa}[\eps_{\nu\alpha\vkp\vmu} \eps_{d\beta\vkp\vnu} \Phi_{\vmu}\Phi_{\vnu}]
\ee
Substituting this into the expression (\ref{pr2}) we find
\bea
&&B_{\tau\tau}[B_{\mu d}\eps_{\nu\alpha\vkp\vmu}\eps_{d\beta\vkp\vnu}+B_{\nu\alpha}\eps_{\mu d\vkp\vmu}\eps_{\beta d\vkp\vnu}]\Phi_{\vmu}\Phi_{\vnu}\phi_\mu\phi_\nu\nonumber\\&=&B_{\tau\tau}[\eps_{\mu\vtau\vrho}\eps_{d\vtau\vsig}\eps_{\nu\alpha\vkp\vmu}\eps_{d\beta\vkp\vnu}-\eps_{\mu\vtau\vrho}\eps_{\alpha\vtau\vsig}\eps_{\nu d\vkp\vmu}\eps_{d \beta\vkp\vnu}]\Phi_{\vmu}\Phi_{\vnu}\Phi_{\vrho}\Phi_{\vsig}\phi_\mu\phi_\nu
\eea
Now, using (\ref{epid}) to move subscript $d$,
\bea
\eps_{d\vtau\vsig}\eps_{\nu\alpha\vkp\vmu}-\eps_{\alpha\vtau\vsig}\eps_{\nu d\vkp\vmu}=\eps_{\nu\vtau\vsig}\eps_{d\alpha\vkp\vmu}+\eps_{\kappa_2\vtau\vsig}\eps_{\nu\alpha d \kappa_3\ldots\kappa_n\vmu}+\ldots+\eps_{\kappa_r\vtau\vsig}\eps_{\nu\alpha\kappa_2\ldots\kappa_{r-1}d\vmu}\nonumber\\+\eps_{\mu_2\vtau\vsig}\eps_{\nu\alpha\vkpp d \mu_3\ldots\mu_n}+\ldots+\eps_{\mu_n\vtau\vsig}\eps_{\nu\alpha\vkpp \mu_2\ldots\mu_{n-1}d}
\eea
As  before the last terms will vanish due to symmetry considerations. For the middle terms, by relabelling and using antisymmetry
\be
\eps_{\kappa_i\vtau\vsig}\eps_{\nu\alpha\kappa_2\ldots d\ldots\kappa_r\vmu}\eps_{d \beta\kappa_2\ldots\kappa_i\ldots\kappa_r\vnu}=\eps_{\kappa_2\vtau\vsig}\eps_{\nu\alpha d\vkpp\vmu}\eps_{d \beta\vkp\vnu}
\ee
There are $(r-1)=(d-n-1)$ of these terms. We now have
\be
\frac{r}{r+1}B_{\tau\tau}[\eps_{\mu\vtau\vrho}\eps_{\nu\vtau\vsig}\eps_{d\alpha\vkp\vmu}\eps_{d\beta\vkp\vnu}+(r-1)\eps_{\mu\vtau\vrho}\eps_{\kappa_2\vtau\vsig}\eps_{\nu\alpha d\vkpp\vmu}\eps_{d \beta\vkp\vnu}]\Phi_{\vmu}\Phi_{\vnu}\Phi_{\vrho}\Phi_{\vsig}\phi_\mu\phi_\nu
\ee
Again, rewriting the epsilons, this time moving subscript $\kappa_2$,
\bea
\eps_{\mu\vtau\vrho}\eps_{d \beta\vkp\vnu}=\eps_{\kappa_2\vtau\vrho}\eps_{d \beta\mu\vkpp\vnu}+\eps_{\mu\kappa_2\tau_2\ldots\tau_r\vrho}\eps_{d \beta\tau_1\vkpp\vnu}+\ldots+\eps_{\mu\tau_1\ldots\tau_{r-1}\kappa_2\vrho}\eps_{d \beta\tau_r\vkpp\vnu}+\nonumber\\\ldots+\eps_{\mu\vtau\kappa_2\rho_3\ldots\rho_n}\eps_{d\beta\rho_2\vkpp\vnu}+\ldots+\eps_{\mu\vtau\rho_2\ldots\rho_{n-1}\kappa_2}\eps_{d\beta\rho_n\vkpp\vnu}
\eea
And again, by relabelling
\be
\eps_{\mu\tau_1\ldots\kappa_2\ldots\tau_r\vrho}\eps_{d \beta\tau_i\vkpp\vnu}\eps_{\kappa_2\vtau\vsig}=-\eps_{\mu\vtau\vrho}\eps_{d\beta\vkp\vnu}\eps_{\kappa_2\vtau\vsig}
\ee
There are $r=d-n$ of these terms. 
So our expression is now
\bea
\frac{r}{r+1}B_{\tau\tau}[\eps_{\mu\vtau\vrho}\eps_{\nu\vtau\vsig}\eps_{d\alpha\vkp\vmu}\eps_{d\beta\vkp\vnu}+\frac{r-1}{r+1}\eps_{\kappa_2\vtau\vrho}\eps_{\kappa_2\vtau\vsig}\eps_{\nu\alpha d\vkpp\vmu}\eps_{d \beta\vkpp\vnu}]\Phi_{\vmu}\Phi_{\vnu}\Phi_{\vrho}\Phi_{\vsig}\phi_\mu\phi_\nu\phi_{\alpha\beta}\nonumber
\eea
\bea
&&=\frac{r}{r+1}B_{\tau\tau}[\eps_{d\alpha\vkp\vmu}\eps_{d\beta\vkp\vnu}\Phi_{\vmu}\Phi_{\vnu}r! {\cal L}-\frac{r-1}{r+1}B_{\kappa\kappa}J_{d\alpha\vkpp} J_{d\beta\vkpp}]\phi_{\alpha\beta}\nonumber\\
&&=\frac{r r!}{r+1}\left(\frac{\pd J_{\vmu}}{\pd \phi_\tau}\frac{\pd J_{\vmu}}{\pd \phi_\tau}\right)\left[\left(\frac{\pd J_{d\vkp}}{\pd \phi_\alpha}\frac{\pd J_{d\vkp}}{\pd \phi_\beta}\right) {\cal L}-\frac{r-1}{(r+1)!}\left(\frac{\pd J_{\vnu}}{\pd \phi_\kappa}\frac{\pd J_{\vnu}}{\pd \phi_\kappa}\right) J_{d\alpha\vkpp}J_{d\beta\vkpp} \right]\phi_{\alpha\beta}\nonumber\\
\eea
A very similar calculation can be carried out to rewrite the coefficients of $\phi^j_{\alpha\beta}$ $(j\neq 1)$  from (\ref{pr1}). These become, for $\phi^j=\psi$, say.
\be
\frac{r r!}{r+1}\left(\frac{\pd J_{\vmu}}{\pd \phi_\tau}\frac{\pd J_{\vmu}}{\pd \psi_\tau}\right)\left[\left(\frac{\pd J_{d\vkp}}{\pd \psi_\alpha}\frac{\pd J_{d\vkp}}{\pd \psi_\beta}\right) {\cal L}-\frac{r-1}{(r+1)!}\left(\frac{\pd J_{\vnu}}{\pd \psi_\kappa}\frac{\pd J_{\vnu}}{\pd \psi_\kappa}\right) J_{d\alpha\vkpp}J_{d\beta\vkpp} \right] \psi_{\alpha\beta}
\ee
When the condition that the Lagrangian vanishes is put into the equations of motion, they can just be written as
\be
J_{d\alpha\vkpp} J_{d\beta\vkpp} \phi^i_{\alpha\beta}=0\label{ans}
\ee
as required.
Comparing (\ref{ans}) with (\ref{eomsq}), these are the equations of motion for the Lagrangian involving a square root (\ref{lagsq}) in $(d-1)$ dimensions.

\section{Conclusion}
It has been proved that the equations of motion for the companion Lagrangian without a square root when subject to some constraints are equivalent to the equations of motion for the companion Lagrangian with a square root in one less dimension.

\section*{Acknowledgements}
I am grateful to David Fairlie for useful discussions and to EPSRC for a postgraduate research award.


\begin{thebibliography}{9}
\bibitem{me1} L. M. Baker, D. B. Fairlie, {\bf hep-th/9908157} (to appear in J. Math. Phys., July 2000).
\bibitem{me2} L. M. Baker, D. B. Fairlie, {\bf hep-th/0003048}.
 \end{thebibliography}
\end{document}